\documentclass[reqno,a4paper]{amsart}
\usepackage{amssymb,stmaryrd}
\usepackage{amsfonts}
\usepackage{amstext}
\usepackage{algorithmic}
\usepackage{algorithm}
\usepackage{color}
\usepackage{graphicx}
\usepackage{epstopdf}
\usepackage[all]{xy}
\usepackage{MnSymbol}
\numberwithin{equation}{section}
\newtheorem{theorem}{Theorem}[section]
\newtheorem{lemma}[theorem]{Lemma}

\theoremstyle{definition}

\newtheorem{example}[theorem]{Example}

\theoremstyle{remark}

\newcommand{\R}{{\mathbb{R}}}

\newcommand{\<}{{\langle}}

\renewcommand{\>}{{\rangle}}

\newcommand{\CO}{{\mathcal{O}}}

\newcommand{\CD}{{\mathcal{D}}}

\newcommand{\extd}{{\rm d}}
\newcommand{\del}{{\partial}}
\newcommand{\eps}{\epsilon}
\renewcommand{\imath}{\mathrm{i}}

\allowdisplaybreaks
\begin{document}

\title{Klein-Gordon flow on FLRW spacetimes}



\thanks{
{\it Authors to whom correspondence should be addressed:} e.j.beggs@swansea.ac.uk  and s.majid@qmul.ac.uk}
\date{Ver 1.1}


 \maketitle 
 
 \begin{center}Edwin Beggs$^a$ and Shahn Majid$^b$\\ \ \\
$^a$ Department of Mathematics, Bay Campus, Swansea University, SA1 8EN, UK\\
$^b$ Queen Mary University of London, 
School of Mathematical Sciences, London E1 4NS, UK
\end{center}
\begin{abstract}  
We study a new approach to generally covariant quantum mechanics applied in the case of an FLRW cosmological background. For positive spatial curvature we find a discrete series of solutions of the Klein-Gordon equation that can reasonably be called gravitationally bound `cosmological atom' states. For all cases of curvature,  these modes, as well as more conventional atomic spatial modes bound by an external potential, extend to solutions of the Klein-Gordon equations viewed as stationary modes of Klein-Gordon quantum mechanics where wavefunctions are over spacetime and evolution is with respect to an external `geodesic time' parameter $s$.  For general nonstationary states with fixed spatial eigenvector, the theory reduces to a novel 1-dimensional quantum system on the time $t$ axis with potential $1/a(t)^2$, where $a(t)$ is the Friedmann expansion factor. Its behaviour, and hence the evolution of spatial states, changes critically when the Hubble constant exceeds $2/3$ of the particle mass, as typically occurs during inflation.  We also find washout of the evolution of spatial observables at late times and a backward-traveling reflected mode generated when the value of $H$ transitions to a larger value. \end{abstract}

\section{Introduction}

Recently\cite{BegMa:gen}, we have proposed a generally covariant formulation of quantum mechanics in which the role of the Heisenberg algebra is played by the algebra $\CD(M)$ of differential operators on a spacetime manifold $M$. Although coming out of a notion of quantum geodesics\cite{Beg:geo,BegMa:geo,BegMa:cur, BegMa:gra, LiuMa} in noncommutative geometry applied to this algebra, the end product is remarkably simple and can be computed for any spacetime without knowing any noncommutative geometry. In any local coordinate chart it amounts to a pair of equations
\begin{align}\label{heis1}
\mu {\extd x^\mu\over\extd s}&= g^{\mu\nu} p_\nu - {\lambda\over 2} \Gamma^\mu,\\ \label{heis2}
\mu {\extd p_\mu\over\extd s}&= \Gamma^\nu{}_{\mu\sigma}g^{\sigma\rho}(p_\nu p_\rho-\lambda\Gamma^\tau{}_{\nu\rho}p_\tau)+{\lambda\over 2}g^{\alpha\beta} \, \Gamma^\nu{}_{\beta\alpha,\mu}\,p_\nu, 
 \end{align}
 for operators $x^\mu, p_\nu$ obeying $[p_\nu,x^\mu]=\lambda\delta^\mu_\nu$, where $\lambda=-\imath\hbar$, $g$ is the metric and $\Gamma^\nu{}_{\mu\sigma}$, $\Gamma^\mu=\Gamma^\mu{}_{\alpha\beta}g^{\alpha\beta}$ are the Christoffel symbols of the Levi-Civita connection and its contraction. The real parameter $\mu$ plays a role of mass (and has mass dimension).  We have omitted an optional external potential $V$. This is a first order phase space formulation of the geodesic equations in an operator Heisenberg picture. The corresponding Schr\"odinger picture turns out to be simply 
 \begin{equation} \label{KGQM}
 -\imath {\del\psi\over\del s}= {\hbar \over 2\mu} \square \psi
 \end{equation} 
 for $s$-dependent wave functions $\psi\in L^2(M)$ extended over spacetime, where $\square$ is the Klein-Gordon (KG) operator.  We will refer to this the Klein-Gordon flow or Klein-Gordon quantum mechanics (KGQM). The external time parameter $s$ makes sense in the classical limit of the Heisenberg picture as  proper time, but how we think of it in the quantum case and in (\ref{KGQM}) is by analogy with the case where $M$ is space and $s$ is the external time of someone viewing geodesics in $M$.  First, the reader should imagine a dust of particles each moving on geodesics an amount $s$ when parameterised by arclength/proper time. This implies that the densities evolve with $s$. We then replace the flow of a density $\rho$ of such particles by the flow of a wave function $\psi$ such that $\rho=|\psi|^2$.  At the density level, there are also similarities with optimal transport\cite{LotVel} and there could be applications to relativistic fluid dynamics\cite{Olt}, but when we work with wave functions the theory acquires a  different and more quantum-mechanics like character. Next,  when $M$ is spacetime, we can still can think of $|\psi|^2$ as the probability density for a hypothetical external `observer' to find a particle in a region of spacetime. If we are happy in GR to watch a single particle and let it evolve on a geodesic by its own proper time $s$, we should be happy to have evolution for a quantum particle also with respect to some kind of proper time $s$ even if its interpretation is now less clear. At the density level, we can still set all particles off on geodesics and let each evolve by $s$, which amounts to a kind of collective proper time for the model. Its physical interpretation in the wave function case then proceeds by analogy. We believe that the full physical meaning of $s$ will firm up over time through examples and applications, for which the  present work is a start. Till then, one can at the very least consider KGQM in (\ref{KGQM}) for evolution with respect to $s$ as a mathematical tool to explore the physics in a new way. Another point of view is that, at the phase space level, the evolution (\ref{heis1})-(\ref{heis2}) is not far from other flows in the literature, such as in \cite{Cha}.
  
In practical terms, the geometry of the relevant quantum geodesic actually takes place  on $\CD(M)$ viewed as a quantum phase space, but this algebra is represented on $L^2(M)$ so there is an induced quantum geodesic `Klein-Gordon flow' on this, which comes out as (\ref{KGQM}). The original motivation for quantum geodesics was in the context of the {\em quantum spacetime hypothesis} that spacetime is better modelled as noncommutative due to quantum gravity effects\cite{Ma:pla,DFR,Hoo,MaRue, BegMa} and leading in turn to baby models of quantum gravity and applications to the problem of the  cosmological constant\cite{BliMa}. In the present work, however, spacetime is classical and we have the usual tools of General Relativity (GR) there, it is the phase space which is quantum. Also note that in KGQM we are not only interested in specific-mass on-shell Klein-Gordon field. Such fields  play a role which in ordinary quantum mechanics would be that of energy eigenfunctions or stationary `evolution eigenstates' for (\ref{KGQM}). A general initial wave function $\psi$ at $s=0$ can be viewed as a linear combination of varying mass KG solutions and then evolved in $s$ with each stationary state evolving with a phase. Even if the reader is only interested in solutions of the KG equations for a given mass, this puts them into a slightly wider context which we will still see is useful. 
  
The   equations (\ref{heis1})--(\ref{KGQM})  were studied in the case of a black hole background\cite{BegMa:gen}, including  a numerical look at an initial real Gaussian `bump' wave function falling into the black hole and in the process replaced by waves generated at (just above) the horizon. The entropy of the density $\rho$ increased as this happened. Also solved was a hydrogen-like gravatom with the black hole in the role of the nucleus, and it was found that the energy is not quantised and that wave functions have a fractal banding approaching the horizon. In the present work, we do similar calculations for an FLRW cosmological background. Section~\ref{secFLRW} recalls the metric and wave operator.  Section~\ref{secSch}  studies KGQM in this background and our results include particular KG solutions/stationary states for KGQM that could be called `cosmological atoms' for a general Friedmann expansion function $a(t)$. Section~\ref{secHeis} computes the operator geodesic equations and some resulting Ehrenfest theorems for the expected values. Unlike the black hole case, where the metric is static but the time component is radially dependent, the situation for the FLRW case is reversed with the interesting result that, while it can again be solved by separation of variables (as we shall see), the natural special case is not pseudo-quantum mechanics in which the spatial factor evolves under $s$ in a quantum mechanics-like manner, but what we call `temporal quantum mechanics' in which the $t$-dependent factor evolves under $s$ and $a(t)$ appears as an arbitrary potential. 

Section~\ref{secexp} completes the picture by showing how our methods also allow the interpretation of solutions of the  Klein-Gordon equation (such as the cosmological atom ones) as quantum mechanics on space with respect to evolution under $t$. We identify three significant effects in this context from high Hubble constant $H$, such as typically occurs in models of inflation. The key difference is that whereas in flat spacetime quantum mechanics, a spatial eigenmode with energy $E_\nu$ evolves with a phase $e^{-\imath {m\over\hbar  }t } e^{-\imath{E_\nu\over\hbar}t}$ (the rest mass factor usually being suppressed  but present in the KG point of view), this is now replaced by $F_\omega(t)$ solving (\ref{TQMi}) as $s$-independent stationary modes for temporal quantum mechanics. Solutions of this  are very far from a simple phase factor when $H>0$. Since our first arXiv version, we learned of a previous work \cite{Fer} which already  noted the relevant solutions $F^\pm_\omega(t)$ for the case of $H$ constant and the consequent polar coordinate solutions to the KG equations for this case and flat space $\kappa=0$. As explained in Section~\ref{secflat}, however, the familiar Bessel function spatial modes for $\kappa=0$ are not individually normalisable and  hence not like bound states in our interpretation (for which we will  need $\kappa>0$). When $a(t)$ is not constant, as happens during a period of inflation, one is forced to have both $F^\pm_\omega$ modes at early time, which takes us beyond regular quantum mechanics but still allows us to $t$-evolve any eigenfunction of the spatial Hamiltonian (including with a potential, such as a nuclear atomic mode) as a solution of the Klein-Gordon equation. 

Section~\ref{secconc} provides some concluding remarks about directions for further work. We use units where $c=1$ and adopt the usual conventions of GR with $-+++$ signature. Numerical plots were obtained using MATHEMATICA.

\section{Recap of the FLRW  metric}\label{secFLRW}

The metric is 
\begin{align*}
\extd s^2=-\extd t^2+a(t)^2\Big(\frac{\extd r^2}{1-\kappa\, r^2}+ r^2 ( \extd\theta^2+\sin^2(\theta)\extd\phi^2)\Big)
\end{align*}
using standard conventions\cite{Car}, where $\kappa$ is a real curvature parameter of inverse area dimensions and $a^2(t)$ is a positive dimensionless scale factor. When $\kappa>0$ we have $r\in (0,\sqrt{\kappa})$ and the spatial geometry is that of a 3-sphere (more precisely, an open patch covering essentially half of one). When $\kappa<0$ the spatial geometry is hyperbolic and $r\in(0,\infty)$. The only non vanishing Christoffel symbols up to $\Gamma^i{}_{jk}=\Gamma^i{}_{kj}$ are
\begin{align*}
&\Gamma^t{}_{rr} = \frac{a\,\dot a}{1-\kappa\, r^2},\quad  \Gamma^t{}_{\theta\theta} =a\,\dot a\, r^2,\quad 
\Gamma^t{}_{\phi\phi} =a\,\dot a\, r^2\,\sin^2\!\theta, \cr
&\Gamma^r{}_{rr} = \frac{\kappa\, r}{1-\kappa\, r^2},\quad 
\Gamma^r{}_{tr} = \frac{\dot a}{a},\quad 
\Gamma^r{}_{\theta\theta} =-r(1-\kappa\,r^2),\quad 
\Gamma^r{}_{\phi\phi} = - r(1-\kappa\,r^2)\,\sin^2\!\theta, \cr
& \Gamma^\theta{}_{t\theta} = \frac{\dot a}{a},\quad  \Gamma^\theta{}_{r\theta} = \frac1r,\quad 
\Gamma^\theta{}_{\phi\phi} = -\sin\theta\, \cos\theta, 
\cr
& \Gamma^\phi{}_{t\phi} = \frac{\dot a}{a},\quad \Gamma^\phi{}_{r\phi} = \frac1r\ ,\ 
\Gamma^\phi{}_{\theta\phi} =\cot\theta.
\end{align*}
(Note here dot is the $t$ detivative.) From these we find
\[ \Gamma^t=3{\dot a\over a},\quad \Gamma^r={3 \kappa r^2-2\over a^2 r},\quad \Gamma^\theta=-{\cot\theta\over a^2 r^2},\quad \Gamma^\phi=0.\]

The only nonzero components of the Ricci tensor are diagonal:
\begin{align*}
&R_{tt} = -3\,\frac{\ddot a}{a},\quad 
R_{rr} = \frac{ a\,\ddot a +2\,\dot a^2 +2\,\kappa   }{1-\kappa\, r^2}\ ,\cr
& R_{\theta\theta} = r^2(a\,\ddot a +2\,\dot a^2 +2\,\kappa ),\quad 
R_{\phi\phi} = r^2(a\,\ddot a +2\,\dot a^2 +2\,\kappa )\ \sin^2\theta\ .
\end{align*}
The Ricci scalar is
\[
R=\frac{6}{a^2}\, (a\,\ddot a + \dot a^2+\kappa)\ .
\]

In the FLRW model, we suppose a stress tensor of the form 
\[ T_{00}=\rho,\quad T_{0i}=T_{i0}=0,\quad T_{ij}=g_{ij}p\]
for density and pressure functions $\rho,p$. Typically one chooses an equation of state $p=w\rho$ for a constant $w$ depending
on the type of contribution, namely $w=0$ for dust, $w=1/3$ for radiation and $w=-1$ for vacuum energy. Even if $w$ is not constant,  the conservation of energy equation  $\nabla_\mu T^\mu{}_0=0$ can be written
\[ {\dot\rho\over\rho}= -3(1+w){\dot a\over a}\]
which in the constant $w$ case means $\rho\propto a^{-3(1+w)}$. Then the Einstein equations become the Friedmann equations
\[ ({\dot a\over a})^2={8\pi G\over 3}\rho-{ \kappa\over a^2},\quad  {\ddot a\over a}=-{4\pi G\over 3}(\rho+3p).\]
The static case here has $a^2= {3\kappa \over 8\pi G\rho}$ and $w=-1/3$ but is not interesting since after rescaling of $t$, it amounts to flat spacetime $\R^{1,3}$. We exclude this case. The  next simplest choice is $a(t)=a_0e^{H t}$ for an actually constant Hubble parameter.  Then the Friedmann equations dictate the pressure and density as
\[ 8\pi G \rho= 3 H^2+{3 \kappa \over a_0^2}e^{-2H t},\quad 8\pi G p= -3 H^2- {\kappa \over a_0^2}e^{-2H t}\]
which has $w\sim -1$ at large $t$. Our main results are for general $a(t)$ but we will use this example for illustrative purposes.

\section{KGQM in an FLRW background by separation of variables}\label{secSch}

The wave operator in the general FLRW background  is 
\[ \square=- \Delta_t + {1\over a^2}\Delta,\quad \Delta_t:={\del^2\over\del t^2}+3 {\dot a\over a}{\del\over\del t},\]
where
\[ \Delta=(1-\kappa r^2){\del^2\over\del r^2}+ \big({2-3\kappa r^2\over r}\big){\del\over\del r}+ {1\over r^2}\del_{sph}^2;\quad \del_{sph}^2={\del^2\over\del\theta^2}+{1\over \sin^2(\theta)}{\del^2\over\del\phi^2}+\cot(\theta){\del\over\del\theta}.\]

Our approach to the KGQM equation (\ref{KGQM}) will be to look for modes that have a separation of variables form
\begin{equation}\label{sepvar}    F(t) \psi(r,\theta,\phi), \end{equation}
which then evolve in $s$. This was the approach in our previous work\cite{BegMa:gen},  where for a static metric we factored out $F(t)=e^{{p_t\over\lambda} t}$ for  a real constant $p_t$ and allowed $\psi=\psi(s,r,\theta,\phi)$ to depend on $s$. This $F(t)$ was an eigenmode for $\del_t^2$ and factoring it out gave us something which we called pseudo-QM as it resembles ordinary quantum mechanics with wave functions over space (but evolution time $s$). A general input state over spacetime could then be Fourier transformed in the $t$ variable and the spatial factor of each mode evolved in $s$. By contrast, in the FLRW case the structure of the metric is opposite and hence, while we make the same factorisation, we proceed oppositely and let $F=F(s,t)$  depend on $s$ and fix an eigenstate $\psi_{\nu}$ of eigenvalue $-\nu$ of $\Delta$. A general initial state over spacetime can be expanded in terms of these and the $F$ factor of each of these evolved, according to
\begin{equation}\label{TQM} -\imath {\del \over\del s} F(s,t)=-{\hbar\over 2\mu}(\Delta_t + {\nu\over a(t)^2})F(s,t)\end{equation}
which {\em looks a lot like 1-dimensional quantum mechanics with potential function $1/a^2$}  but with evolution time $s$ and wavefunctions spread over the time axis in place of space. This {\em temporal QM} replaces the role in the FLRW case of pseudo-QM for static metrics\cite{BegMa:gen}. The meaning of $F(s,t)$ is the amplitude for the hypothetical external `observer' at time $s$ to see at time $t$ the fixed eigenfunction $\psi_\nu$ for the form of the spatial distribution. Moreover, we should be careful to use the correct measure. The pseudo-Riemannian measure from $\sqrt{-\det(g)}$ is
\[ a^3(t)\extd t\,  {r^2 \extd r\over \sqrt{1-\kappa r^2}}\sin(\theta) \extd\theta\extd\phi\]
which we see factorises into a part which we will use for the Hilbert space on which $\Delta$ acts and,  relevant now, $a^3\extd t$ for the measure in temporal QM. So 
\[ \<F|F\>=\int  |F|^2 a^3\extd t,\]
where the endpoints should be chosen depending on $a$. Note that the hypothetical external `observer' by definition sees all of spacetime, so the full range $(-\infty,\infty)$ could be a natural option. It should be remembered, however, that this point of view is a mathematical tool and not a physical observer. 

\begin{lemma} $\Delta_t$ is essentially self-adjoint with respect to the $a^3\extd t$ measure on the space of fields for which $[\bar G(\del_t F) a^3]=0$ across the endpoints (for example, any mix of Neumann and Dirichlet conditions at the two limits). 
\end{lemma}
{\bf Proof}
\begin{align*} \int\bar G&(\del_t + 3 {\dot a\over a})(\del_t F) a^3\extd t = \int \bar G (\del_t^2 F) a^3\extd t + \int \bar G (\del_t a^3)\del_t F \extd t\\
&=[\bar G(\del_t F) a^3]-\int (\del_t\bar G)(\del_t F)a^3\extd t\end{align*}
which we then reverse by a similar calculation on the other side to obtain $\int ((\del_t + 3 {\dot a\over a})(\del_t G)F a^3\extd t$. \hfill $\square$

There is no issue with the potential term  $\nu/a^2$ as this just acts by multiplication. We see that on factorisable wave functions, KGQM indeed factorises for each eigenmode of $\Delta$ as `temporal quantum mechanics', provided we use this measure. By a similar calculation, one can check that the radial part of $\Delta$ is similarly hermitian with respect to the ${r^2 \extd r\over\sqrt{1-\kappa r^2}}$ measure on radial functions provided we have boundary conditions on fields $\psi(r),\phi(r)$ so as to be able to drop $[\bar\phi(\del_r\psi) r^2\sqrt{1-\kappa r^2}]$ across the relevant limits. That it also works for the angular dependence with the full spatial measure is the same as for flat spacetime. 

\begin{figure}
\[\includegraphics[scale=0.7]{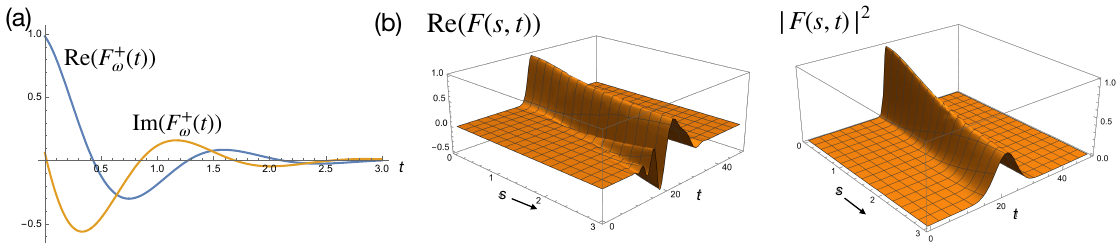}\]
\caption{(a) Stationary state $F^+_\omega(t)$ in the oscillatory regime, shown for $H=1, \omega= 4$ and (b) temporal QM evolution under $s$ of an initial Gaussian $F(0,t)$ centred at $t=25$, shown for $H=0.1$. Both parts are for $a_0=1, \mu=0.5, \nu=10$. \label{figF}}
\end{figure}

\begin{example}\rm\label{exH} For $a(t)=a_0 e^{H t}$ corresponding to a constant Hubble parameter $H>0$, the temporal quantum mechanics equation (\ref{TQM}) has stationary   eigenstates $F^\pm_\omega(t)$ for eigenvalue $-\omega^2$ of $\Delta_t + {\nu\over a(t)^2}$, given along with their evolution by
\[ F^\pm_\omega(t)=c_\pm e^{-\frac{3}{2} H t} J_{\pm\sqrt{{9\over 4}- { \omega^2\over H^2}}}\left(\frac{ \sqrt{\nu}}{ H a(t)}\right),\quad F(s,t)=e^{\imath{\omega^2\hbar\over 2\mu}s}F^\pm_\omega(t)\]
in terms of Bessel $J$ functions and some complex normalisations $c_\pm$, cf Ref. \cite{Fer} where this eigenvalue equation was similarly noted and solved. We assume $\nu\ne0$ and normalise with 
\[ c_\pm= ({\sqrt{\nu}\over 2 a_0 H})^{\mp \sqrt{{9\over 4}- { \omega^2\over H^2}}}\Gamma(1\pm \sqrt{{9\over 4}- { \omega^2\over H^2}}) \]
in terms of a Gamma function. There are two distinct regimes. (i) For $|\omega|>{3 H\over 2}$, we take out $\imath$ from the square root. The $F^\pm_\omega$ are complex oscillatory  with $F^-_\omega$ the complex conjugate of $F^+_\omega$ and $|F^+_\omega|$ decaying exponentially and given asymptotically by 
\begin{equation}\label{oscappx} F^\pm_\omega  \sim  e^{-{3\over 2}H t}e^{\pm\imath\omega' t},\quad \omega'=\sqrt{\omega^2-{9\over 4}H^2}\end{equation}
for  $t>> t_c$ independently of $\nu$, where
\begin{equation}\label{tc} t_c ={1\over H} \ln( {\sqrt{|\nu|}\over  a_0 \omega}). \end{equation} 
This is because  then $a^2(t)\omega^2>>\nu$ so that $\nu$ can be ignored in the eigenfunction equation. A different, small $t,H$, expansion will be given later, in Section~\ref{secexp}. (ii) For other values of $\omega$, the $F^\pm_\omega$ are real and  decay or increase exponentially and are given asymptotically by 
\begin{equation}\label{expappx} F^\pm_\omega \sim e^{-tH({3\over 2} \pm   \sqrt{{9\over 4}- { \omega^2\over H^2}})}\end{equation}
for  $t>>t_c$, independently of $\nu$. 

 From (\ref{oscappx}), it is clear that $F_\omega^\pm$ in the oscillatory regime should be viewed as positive/negative energy solutions  for the operator $-\imath D_t$ in the sense of value $\pm\omega'$ for large $t$, where $D_t:=\del_t+ {3\over 2}{\dot a\over a}=\del_t+ {3\over 2}H$ in our case. Factoring out an $F^+_\omega(t)$ time dependence in the FLRW case thus plays the  role of factoring out $e^{{p_t\over\lambda}t}$ in the black hole case\cite{BegMa:gen}. The $F^\pm_\omega$  modes are not,  however, in an appropriate Hilbert space for the $a^3$ measure. We do not necessarily care about this since, being analogues of plane waves in temporal QM, we don't insist that they are normalisable. However,  by taking linear combinations of the $F^\pm_\omega$ in the oscillatory regime we can find a `sine'  version $F_\omega(t)$ where $F_\omega(0)=0$ and $F'_\omega(\infty)=0$ so that we can work in the half-line $t\in [0,\infty)$ with mixed Dirichlet/Neumann boundary conditions at the two ends. Similarly, there is a `cosine' version with Neumann at both ends. However, the $a^3$ factor in the measure still prevents such modes from being square-integrable with respect to the $a^3\extd t$ measure by cancelling the $e^{-{3\over 2}Ht}$, in keeping with their plane wave character.

On the other hand, any initial $F(0,t)$ can be evolved directly from (\ref{TQM}). For example, the evolution of an initial Gaussian bump $F(0,t)$ for the same Hubble constant $a(t)$ as here is also shown in Figure~\ref{figF}. We see that its density $|F(s,t)|^2$ spreads out as $s$ increases while $F$ develops complex oscillations, as to be expected for this form of PDE. Provided we stay away from the endpoints, as we do,  we remain in the space of fields that are (to a good approximation) zero at the endpoints of $t$. In the example, we took $t\in (0,t_{\rm max})$ where $t_{max}=50$  and we also use this for the integration in computing norms and expectation values. As a check of the numerical integrity, we verified that $\<F|F\>$ is indeed constant in $s$ to within the level of numerical noise (it changes by $\pm 0.004 \%$ over the range here), in keeping with the evolution being unitary with respect to this measure. 
\end{example}

Next, we are particularly interested in stationary `evolution eigenstates', i.e. the $s$-independent  KGQM equation, which just means solving the KG equation for an eigenvalue $E_{KG}$ in place of $m^2$, more precisely modes on spacetime where 
\[ \square =  -{2\mu\over\hbar^2}E_{KG} \]
in the conventions of our previous work\cite{BegMa:gen}. We are interested in the case where $E_{KG}$ is real but not necessarily  positive or zero; we are not looking for massive or massless scalar fields but rather the theory is `off shell' in potentially looking at all eigenvalues. From the above analysis we see that separable solutions as in (\ref{sepvar}) are of the form $F_\omega(t) \psi_\nu$ for $\Delta\psi_\nu=-\nu\psi_\nu$ in the spatial sector and $F_\omega$ solving the {\em time-independent temporal Schr\"odinger equation} 
\begin{equation}\label{TQMi} (\Delta_t+ {\nu \over a^2})F_\omega= - \omega^2 F_\omega\end{equation}
to give a solution of the KG equation/evolution eigenstate with
\begin{equation}\label{mKG}  -2\mu E_{KG}=\hbar^2 \omega^2=m^2_{KG}, \end{equation}
where we also give the corresponding mass for the Klein-Gordon equation. We have seen examples of solutions of this in Example~\ref{exH}. The Klein-Gordon equation on an FLRW background does not appear to have been studied in detail at the level of exact solutions, but we note important decay estimates\cite{Rin} and recent work\cite{Fer}. In particular, the separation of variables approach, which we will combine with methods familiar for the hydrogen atom to determine the $\psi$ eigenstates, will lead to a particular class of solutions of the Klein-Gordon equations for $\kappa>0$ that we call `cosmological atoms'.

\subsection{Polar-separable eigenfunctions of $\Delta$}

Following the usual methods for the hydrogen atom, we separate out the angular degrees of freedom by looking for eigenfunctions of $\Delta$ of the form
\[ \psi(r)Y_l^m(\theta,\phi),\]
where
\[ Y_l^m(\theta,\phi)\propto e^{\imath m\phi}P^m_l(\theta),\quad \del_{sph}^2 Y_l^m=-l(l+1) Y_l^m,\quad {\del\over\del \phi}Y_l^m=\imath m Y_l^m\]
are the standard spherical harmonics for integers $l\ge 0,m$ with  $-l\le m\le l$. In this case the eigenvector equation for $\Delta$ becomes on $\psi(r)$, 
\[ \left((1-\kappa r^2){\del^2\over\del r^2}+ \big({2-3\kappa r^2\over r}\big){\del\over\del r}- {l(l+1)\over r^2} \right)\psi=- \nu  \psi\]
for some real constant $\nu$. This can be solved in terms of ${}_2F_1$ hypergeometric functions with two modes, one of which diverges at $r=0$. Excluding this, there is a unique nonsingular solution for $\kappa\ne 0$,
\[ \psi_{\nu,l}(r)=r^l \, _2F_1\left(\frac{l +1 -\sqrt{ 1+{\nu\over \kappa}}}{2 },\frac{l +1 +\sqrt{ 1+{\nu\over \kappa}}}{2 }, l+\frac{3}{2}, \kappa  r^2\right).\]
This is real-valued for all real values of the parameters and $r\ge 0$, and can also be written in terms of Legendre functions. The $l=0$ modes can be written more simply as 
\[ \psi_{\nu,0}(r)={1\over r\sqrt{\kappa+\nu}} \sin \left( \sqrt{1+\frac{\nu}{\kappa}}  \sin ^{-1}(\sqrt{\kappa}r)\right).\] 

\subsubsection{$\kappa>0$ spherical case.}\label{secatom}  We first look at the spherical case $\kappa>0$ and $r\le 1/\sqrt{\kappa}$. Here $\psi(r)$ is real and bounded over the allowed values of $r$, even at $r=1/\sqrt{\kappa}$. The same cannot be said for $\psi'$ which diverges there for all but a discrete series of $\nu>0$, where it  vanishes. These special values are of the form
\[ \nu=\kappa(n^2-1);\quad n> l,\quad n-l \ {\rm odd}.\]
For example, for $l=0$ we have $n=1,3,5,\cdots$ and for $l=1$ we have $n=2,4,6,\cdots$, etc. These modes correspond a subspace of the $n^2$-dimensional space of  matrix elements of the $n$-dimensional representation of $SU(2)$  as expected from group theory, where $\Delta$ is the action of the quadratic Casimir. Indeed, if we do not impose the restriction that $n-l$ is odd then the number of modes  for a given $n$ would be $\sum_{l=0}^{n-1} (2l+1)=n^2$ where for orbital angular momentum $l$ there are $2l+1$ modes as we vary $-l\le m\le l$ and we can only take up to $l=n-1$ for $n>l$.  

\begin{figure}
\[\includegraphics[scale=.65]{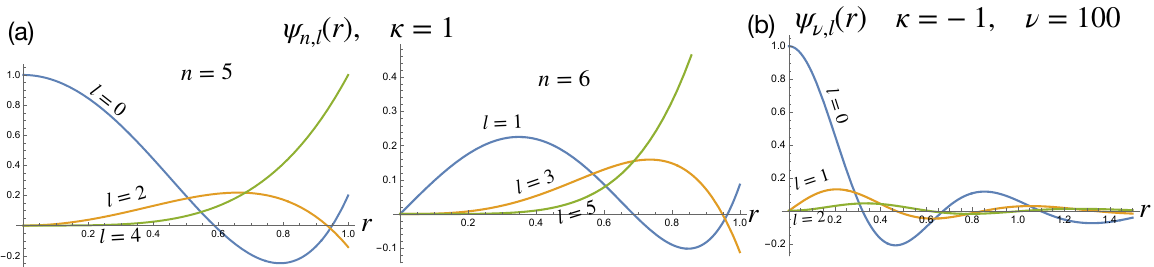}\]
\caption{Radial sector for (a) $\kappa>0$ with `cosmological atom'-like modes $\psi_{n,l}(r)$ for $n=5$ and $n=6$ with allowed values of $l$ in each case. (b) $\kappa<0$  typical form of oscillatory decaying solutions $\psi_{\nu,l}$. \label{fig1}}
\end{figure}

 In summary, we have stationary modes (`bound states')  as some kind of `cosmological atom'. The $\psi_{n,l}$ states (labelled by level $n$ rather than $\nu$)  are plotted in Figure~\ref{fig1} for $n=5$ in one plot and $n=6$ in another. There are $(n-l-1)/2=k$ `quarter-cycles' as measured by the number of zero crossings. These modes are not dissimilar to the radial modes for the black-hole gravatom\cite{BegMa:gen}, but without the fractal aspects there. When multiplied by $F^\pm_\omega$, we obtain a corresponding discrete series of solutions of the KG equations/stationary states in KGQM singled out by this construction. An interpretation of such KG solutions in terms of quantum mechanics with respect to $t$ is deferred to Section~\ref{secexp}. 

\subsubsection{$\kappa=0$ spatially flat case.}\label{secflat} In this case $\Delta$ is the Laplacian of $\R^3$ and the spatial eigenmodes and their eigenvalue are just
\[ \psi_{\vec k}(\vec x)=e^{\imath \vec x\cdot\vec x},\quad \nu=|\vec k|^2,\]
giving solutions of KG equation/stationary states for KGQM of the form 
\[ \psi_{\omega,\vec k}(t,\vec x)=F^\pm_\omega(t)\psi_{\vec k}(\vec x)\]
with mass (\ref{mKG}). However, we can also look for polar-separable eigenfunctions of $\Delta$ where we factor out $Y^m_l$ as above. The radial equation is well-known for all $l$ to give spherical Bessel functions $\psi_{\nu,l}(r)=j_l\left(\sqrt{\nu} r\right)$ if we fix boundary conditions $\psi_{\nu,l}(0)=1$ in order to be nonsingular at $r=0$ (the other Bessel mode is singular there).  We multiply this by $Y^m_l$ with $-l\le m\le l$, to obtain  polar-separable bounded eigenfunctions of $\Delta$ with $\nu$  a continuous parameter. Multiplying by $F^\pm_\omega$ then gives a class of solutions\cite{Fer} of the KG equations/stationary states in KGQM . Note, however, that these spherical Bessel functions are not normalisable over $\R^3$ when the correct $r^2\extd r$ measure is used, so we do not have any bound states. This is just as well as such as state would imply in ordinary QM  a hydrogen atom of charge 0.

\subsubsection{$\kappa<0$ hyperbolic case.}\label{sechyp} This time $0\le r <\infty$ and $\psi_\nu(r)$ and its derivative diverge logarithmically as $r\to \infty$ when $\nu$ is strictly negative, and $\psi_\nu$ is constant for $\nu=0$. For $\nu>0$, both the function and its derivative are bounded and decay as $r\to \infty$,  and are oscillatory for $\nu>-\kappa$. This gives a continuous series of real modes $\psi_{\nu,l}(r)$ for each $l$. Some examples are shown in Figure~\ref{fig1} (c) for the first three $l$. These modes do not, however, appear to be square integrable with respect to the hyperboloid measure ${r^2\over \sqrt{1-\kappa r^2}}\extd r$ due to a log divergence. Multiplying by $F^\pm_\omega$  gives continuous families of solutions of the KG equations/stationary states in KGQM.

\section{Operator geodesic equation in an FLRW background} \label{secHeis}

We first compute the relations from (\ref{heis1}) for the FLRW metric and the  coordinate basis. Inserting the relevant expressions immediately gives
\begin{align}\label{dxds}
\mu {\extd t\over\extd s}&= - p_t - \lambda\, {3\,\dot a\over 2\, a} \ ,\cr
\mu {\extd r\over\extd s}&= {1\over a^2}\left((1-\kappa\, r^2) \,  p_r - \lambda\, {3 \kappa r^2-2\over 2  r}\right), \cr
\mu  {\extd \theta\over\extd s}&={1\over a^2}\left( \frac{1}{ r^2} \, p_\theta + \lambda\, {\cot\theta\over  2 r^2}\right),\cr
\mu {\extd \phi\over\extd s}&= {1\over a^2}\left( \frac{1}{ r^2\,\sin^2(\theta)} \,  p_\phi\right).
 \end{align}

Recall that we are writing our wave functions in the factorised form $F(s,t)\psi_\nu(r,\theta,\phi)$  where $\psi_\nu$ is a fixed eigenvector of the spatial Laplacian $\Delta$ with eigenvalue $-\nu$ (and any wave function will be a (possibly continuous) linear combination of such modes under a spectral decomposition). On such factorised modes,  it is automatic that 
\[{\extd \< \CO(r,\theta,\phi) \>\over\extd s}=0\]
for any operator that does not involve $t, p_t$. This is because the $\int |F(s,t)|^2 a^3 \extd t $ cancels above and below in the calculation of  $\< \CO(r,\theta,\phi) \>$ (this would no longer be true for indecomposable i.e. entangled states between the $t$ and spatial sectors). This observation combined with the Ehrenfest theorem implies 
\[ 0=\<F|a^{-2}|F\>\< \psi_\nu|\CO|\psi_\nu\>\]
for the different spatial $\CO$ on the right and side in (\ref{dxds}). But 
$\<F|a^{-2}|F\>=\int |F|^2 a\extd t\ne 0$ holds for all $s$ (we only need it to hold for some $s$), hence we must have at least in the $\kappa>0$ case 
\[ \int_0^{1\over\sqrt{\kappa}}\extd r \, r^2\sqrt{1-\kappa r^2}\psi_{\nu,l}(r){\del\over\del r} \psi_{\nu,l}(r)= \int_0^{1\over\sqrt{\kappa}} \extd r \, {r^2\over \sqrt{1-\kappa r^2}} {3\kappa r^2-2\over 2r} \psi_{\nu,l}(r)^2\]
\[ \int_0^{2\pi} {P^m_l(\theta)^2\over \sin(\theta)}\extd\theta=0,\quad \int_0^{2\pi}P^m_l(\theta)(\sin(\theta){\del\over\del\theta}+ {\cos(\theta)\over 2})P^m_l(\theta)\extd\theta=0.\]
The latter two identities are less obvious properties of Legendre polynomials/spherical harmonics which can, however, be verified. The first identity must also hold and presumably follows from the radial equation for $\psi_{\nu,l}(r)$ and integration by parts. We have verified it directly for the solutions in Figure~\ref{fig1}(a). The flat and hyperbolic cases do not obey this, however these modes are not normalisable so their analysis is more complicated. 

By contrast, the first of (\ref{dxds}) has cancellation of the spatial integrals and becomes an Ehrenfest identity in the effective temporal QM, 
\begin{equation}\label{dvevtds} \mu {\extd \< t \>\over \extd s}= -\< P_t\>,\quad P_t:=p_t+\lambda {3\over 2}{\dot a\over a}\end{equation}
remembering that $p_t$ acts as $\lambda {\del\over\del t}$. This makes sense if we work with functions $F,G$ such that
\[  [\bar G F a^3]=0\]
accross the limits of integration, for then, using the residual measure $a^3\extd t$ from the metric, 
\begin{align*} \int \bar G (\del_t + {3\dot a\over 2 a})F a^3\extd t&=[\bar GF a^3] -\int  ((\del_t - {3\dot a\over 2 a})\bar G)F a^3\extd t-\int \bar G F 3 \dot a a^2 \extd t\\
&=-\int ((\del_t + {3\dot a\over 2 a})\bar G)F a^3\extd t  \end{align*}
so that $\lambda$ times this operator, i.e. the action of $P_t$, is essentially self adjoint. 

We now compute the content of (\ref{heis2}). Again inserting the form of the FLRW metric but this time with a lot more computation. We first recall the spherical momentum\cite{BegMa:gen}, 
\[
p_{\mathrm{sph}}^2=p_\theta{}^2+\frac1{\sin^2(\theta)}\,p_\phi{}^2 + \lambda\,\cot(\theta)\, p_\theta. 
\]
Then 
\begin{align}\label{dpphids}
\mu {\extd p_\phi\over\extd s} &= \Gamma^\nu{}_{\phi\sigma}g^{\sigma\rho}(p_\nu p_\rho-\lambda\Gamma^\tau{}_{\nu\rho}p_\tau) \cr
&= \Gamma^\phi{}_{\phi\nu}g^{\nu\nu}(p_\phi p_\nu-\lambda\Gamma^\tau{}_{\phi\nu}p_\tau) 
+ \Gamma^\nu{}_{\phi\phi}g^{\phi\phi}(p_\nu p_\phi-\lambda\Gamma^\tau{}_{\nu\phi}p_\tau) \cr
&= \big( \Gamma^\phi{}_{\phi\nu}g^{\nu\nu}
+ \Gamma^\nu{}_{\phi\phi}g^{\phi\phi} \big) (p_\nu p_\phi-\lambda\Gamma^\tau{}_{\nu\phi}p_\tau) \cr
&= \big( \Gamma^\phi{}_{\phi\nu}g^{\nu\nu}
+ \Gamma^\nu{}_{\phi\phi}g^{\phi\phi} \big) (p_\nu p_\phi-\lambda\Gamma^\phi{}_{\nu\phi}p_\phi) =0
 \end{align}
 after some calculation.  Next, 
\begin{align}\label{dpthetads}
\mu {\extd p_\theta\over\extd s}&=\Gamma^\nu{}_{\theta\sigma}g^{\sigma\rho}(p_\nu p_\rho-\lambda\Gamma^\tau{}_{\nu\rho}p_\tau)+  \tfrac{\lambda}{2}  \big( \Gamma^\nu{}_{,\theta}
-   \,g^{\alpha\beta}{}_{,\theta} \, \Gamma^\nu{}_{\beta\alpha} \big)\,p_\nu \cr
&=\Gamma^\theta{}_{\theta\sigma}g^{\sigma\sigma}(p_\theta p_\sigma-\lambda\Gamma^\tau{}_{\theta\sigma}p_\tau)+  
\Gamma^\sigma{}_{\theta\theta}g^{\theta\theta}(p_\sigma p_\theta-\lambda\Gamma^\tau{}_{\sigma\theta}p_\tau) \cr 
&\quad + \Gamma^\phi{}_{\theta\phi}g^{\phi\phi}(p_\phi p_\phi-\lambda\Gamma^\tau{}_{\phi\phi}p_\tau) \cr
&\quad +   \tfrac{\lambda}{2}   \Gamma^\nu{}_{,\theta} \,p_\nu
-   \tfrac{\lambda}{2}   \,g^{\alpha\beta}{}_{,\theta} \, \Gamma^\nu{}_{\beta\alpha} \,p_\nu  \cr
&= \big( \Gamma^\theta{}_{\theta\sigma}g^{\sigma\sigma}  +  
\Gamma^\sigma{}_{\theta\theta}g^{\theta\theta}  \big) (p_\sigma p_\theta-\lambda\Gamma^\tau{}_{\sigma\theta}p_\tau)    \cr 
&\quad + \Gamma^\phi{}_{\theta\phi}g^{\phi\phi}(p_\phi p_\phi-\lambda\Gamma^\tau{}_{\phi\phi}p_\tau) +   \tfrac{\lambda}{2}   \Gamma^\theta{}_{,\theta} \,p_\theta
-   \tfrac{\lambda}{2}   \,g^{\phi\phi}{}_{,\theta} \, \Gamma^\nu{}_{\phi\phi} \,p_\nu  \cr
&= \big( \Gamma^\theta{}_{\theta t}g^{tt}  +  
\Gamma^t{}_{\theta\theta}g^{\theta\theta}  \big) (p_t p_\theta-\lambda\Gamma^\tau{}_{t\theta}p_\tau)    \cr   
&\quad +  \big( \Gamma^\theta{}_{\theta r}g^{rr}  +  
\Gamma^r{}_{\theta\theta}g^{\theta\theta}  \big) (p_r p_\theta-\lambda\Gamma^\tau{}_{r\theta}p_\tau)    \cr     
&\quad + \Gamma^\phi{}_{\theta\phi}g^{\phi\phi}(p_\phi p_\phi-\lambda\Gamma^\tau{}_{\phi\phi}p_\tau) +   \tfrac{\lambda}{2}   \Gamma^\theta{}_{,\theta} \,p_\theta
-   \tfrac{\lambda}{2}   \,g^{\phi\phi}{}_{,\theta} \, \Gamma^\nu{}_{\phi\phi} \,p_\nu  \cr
&=  \frac{\cos(\theta)}{a^2\, r^2\,\sin^3(\theta)} (p_\phi p_\phi-\lambda\Gamma^\tau{}_{\phi\phi}p_\tau) +   \tfrac{\lambda}{2}   \Gamma^\theta{}_{,\theta} \,p_\theta   +\lambda\,  \frac{\cos(\theta)}{a^2\, r^2\,\sin^3(\theta)} \, \Gamma^\nu{}_{\phi\phi} \,p_\nu  \cr
&=  \frac{\cos(\theta)}{a^2\, r^2\,\sin^3(\theta)} \, p_\phi p_\phi  +   \frac{\lambda}{2} 
\,\frac{1}{a^2\,r^2\,\sin^2(\theta)} \,p_\theta.    
 \end{align}
 
 Next, we have 
\begin{align}\label{dprds}
\mu {\extd p_r\over\extd s}&=\Gamma^\nu{}_{r\sigma}g^{\sigma\rho}(p_\nu p_\rho-\lambda\Gamma^\tau{}_{\nu\rho}p_\tau)+  \tfrac{\lambda}{2}  \big( \Gamma^\nu{}_{,r}
-   \,g^{\alpha\beta}{}_{,r} \, \Gamma^\nu{}_{\beta\alpha} \big)\,p_\nu \cr
&=\Gamma^t{}_{rr}g^{rr}(p_t p_r-\lambda\Gamma^\tau{}_{t r}p_\tau)
+ \Gamma^r{}_{rr}g^{rr}(p_r p_r-\lambda\Gamma^\tau{}_{r r}p_\tau) \cr
&\quad + \Gamma^\theta{}_{r\theta}g^{\theta\theta}(p_\theta p_\theta-\lambda\Gamma^\tau{}_{\theta\theta}p_\tau)
+ \Gamma^\phi{}_{r\phi}g^{\phi\phi}(p_\phi p_\phi-\lambda\Gamma^\tau{}_{\phi\phi}p_\tau)  \cr
&\quad + \Gamma^r{}_{rt}g^{tt}(p_r p_t-\lambda\Gamma^\tau{}_{rt}p_\tau)   
 +  \tfrac{\lambda}{2}  \big( \Gamma^\nu{}_{,r}  -   \,g^{\alpha\beta}{}_{,r} \, \Gamma^\nu{}_{\beta\alpha} \big)\,p_\nu \cr
 &= \frac{\kappa\, r}{a^2} \,(p_r p_r-\lambda\Gamma^\tau{}_{r r}p_\tau) \cr
&\quad + r^{-1}\big(g^{\theta\theta}(p_\theta p_\theta-\lambda\Gamma^\tau{}_{\theta\theta}p_\tau)
+ g^{\phi\phi}(p_\phi p_\phi-\lambda\Gamma^\tau{}_{\phi\phi}p_\tau) \big) \cr
&\quad 
 +  \tfrac{\lambda}{2}  \big( \Gamma^\nu{}_{,r}  -   \,g^{\alpha\beta}{}_{,r} \, \Gamma^\nu{}_{\beta\alpha} \big)\,p_\nu \cr
  &= \frac{\kappa\, r}{a^2} \,(p_r p_r-\lambda\Gamma^\tau{}_{r r}p_\tau) 
   + r^{-1}\big(g^{\theta\theta}\, p_\theta p_\theta
+ g^{\phi\phi}\, p_\phi p_\phi \big) \cr
&\quad  +  \tfrac{\lambda}{2}  \big( \Gamma^\nu{}_{,r}  -   \,g^{\alpha\beta}{}_{,r} \, \Gamma^\nu{}_{\beta\alpha} \big)\,p_\nu
 - \lambda\,r^{-1}\big(g^{\theta\theta}\, \Gamma^\nu{}_{\theta\theta}
+ g^{\phi\phi}\, \Gamma^\nu{}_{\phi\phi} \big)p_\nu \cr
  &= \frac{\kappa\, r}{a^2} \,p_r p_r
   + r^{-1}\big(g^{\theta\theta}\, p_\theta p_\theta
+ g^{\phi\phi}\, p_\phi p_\phi \big) \cr
&\quad  +  \tfrac{\lambda}{2}  \big( \Gamma^\nu{}_{,r}  -   \,g^{\alpha\beta}{}_{,r} \, \Gamma^\nu{}_{\beta\alpha} \big)\,p_\nu
 - \lambda\,r^{-1}\big(g^{\theta\theta}\, \Gamma^\nu{}_{\theta\theta}
+ g^{\phi\phi}\, \Gamma^\nu{}_{\phi\phi} \big)p_\nu -  \frac{\kappa\, r}{a^2} \,\lambda\Gamma^\tau{}_{r r}p_\tau \cr
  &= \frac{\kappa\, r}{a^2} \,p_r p_r
   + r^{-1}\big(g^{\theta\theta}\, p_\theta p_\theta
+ g^{\phi\phi}\, p_\phi p_\phi \big) \cr
&\quad  +  \tfrac{\lambda}{2}  \big( \Gamma^t{}_{,r}  -   \,g^{\alpha\beta}{}_{,r} \, \Gamma^t{}_{\beta\alpha} \big)\,p_t
 - \lambda\,r^{-1}\big(g^{\theta\theta}\, \Gamma^t{}_{\theta\theta}
+ g^{\phi\phi}\, \Gamma^t{}_{\phi\phi} \big)p_t -  \frac{\kappa\, r}{a^2} \,\lambda\Gamma^t{}_{r r}p_t \cr
&\quad  +  \tfrac{\lambda}{2}  \big( \Gamma^r{}_{,r}  -   \,g^{\alpha\beta}{}_{,r} \, \Gamma^r{}_{\beta\alpha} \big)\,p_r
 - \lambda\,r^{-1}\big(g^{\theta\theta}\, \Gamma^r{}_{\theta\theta}
+ g^{\phi\phi}\, \Gamma^r{}_{\phi\phi} \big)p_r -  \frac{\kappa\, r}{a^2} \,\lambda\Gamma^r{}_{r r}p_r \cr
&\quad  +  \tfrac{\lambda}{2}  \big( \Gamma^\theta{}_{,r}  -   \,g^{\alpha\beta}{}_{,r} \, \Gamma^\theta{}_{\beta\alpha} \big)\,p_\theta
 - \lambda\,r^{-1}\big(g^{\theta\theta}\, \Gamma^\theta{}_{\theta\theta}
+ g^{\phi\phi}\, \Gamma^\theta{}_{\phi\phi} \big)p_\theta -  \frac{\kappa\, r}{a^2} \,\lambda\Gamma^\theta{}_{r r}p_\theta \cr
&\quad  +  \tfrac{\lambda}{2}  \big( \Gamma^\phi{}_{,r}  -   \,g^{\alpha\beta}{}_{,r} \, \Gamma^\phi{}_{\beta\alpha} \big)\,p_\phi
 - \lambda\,r^{-1}\big(g^{\theta\theta}\, \Gamma^\phi{}_{\theta\theta}
+ g^{\phi\phi}\, \Gamma^\phi{}_{\phi\phi} \big)p_\phi -  \frac{\kappa\, r}{a^2} \,\lambda\Gamma^\phi{}_{r r}p_\phi \cr
  &= \frac{\kappa\, r}{a^2} \,p_r p_r
   + r^{-1}\big(g^{\theta\theta}\, p_\theta p_\theta
+ g^{\phi\phi}\, p_\phi p_\phi \big) \cr
&\quad  +  \tfrac{\lambda}{2}  \big(  -   \,g^{\alpha\beta}{}_{,r} \, \Gamma^t{}_{\beta\alpha} \big)\,p_t
 - \lambda\,r^{-1} \frac{2\,\dot a}{a}\,p_t -  \frac{\kappa\, r}{a^2} \,\lambda\,  \frac{a\,\dot a}{1-\kappa\, r^2}\, p_t \cr
&\quad  +  \tfrac{\lambda}{2}  \big( \Gamma^r{}_{,r}  -   \,g^{\alpha\beta}{}_{,r} \, \Gamma^r{}_{\beta\alpha} \big)\,p_r
 + \lambda\, \frac{2(1-\kappa\,r^2)}{a^2\,r^2} \, p_r -  \frac{\kappa\, r}{a^2} \,\lambda \, \frac{\kappa\, r}{1-\kappa\, r^2}\, p_r \cr
&\quad  +  \tfrac{\lambda}{2}  \big( \Gamma^\theta{}_{,r}  -   \,g^{\phi\phi}{}_{,r} \, \Gamma^\theta{}_{\phi\phi} \big)\,p_\theta
 - \lambda\,r^{-1}\big( g^{\phi\phi}\, \Gamma^\theta{}_{\phi\phi} \big)p_\theta  \cr
   &= \frac{\kappa\, r}{a^2} \,p_r p_r
   + r^{-1}\big(g^{\theta\theta}\, p_\theta p_\theta
+ g^{\phi\phi}\, p_\phi p_\phi \big)  +  \tfrac{\lambda}{2}  \, \Gamma^r{}_{,r}  \,p_r      +  \tfrac{\lambda}{2}  \,\Gamma^\theta{}_{,r} \,p_\theta \cr
   &= \frac{\kappa\, r}{a^2} \,p_r p_r
   + r^{-1}\big(g^{\theta\theta}\, p_\theta p_\theta
+ g^{\phi\phi}\, p_\phi p_\phi \big)  +  \frac{\lambda}{2}  \,{3 \kappa r^2+2\over a^2 r^2} \,p_r      +  \lambda  \,{\cot\theta\over a^2 r^3} \,p_\theta \cr
   &= \frac{\kappa\, r}{a^2} \,p_r p_r
+  \frac{\lambda}{2}  \,{3 \kappa r^2+2\over a^2 r^2} \,p_r     + \frac{1}{a^2\, r^3} \, p_{\mathrm{sph}}^2 .
 \end{align}
Finally, using $\Gamma^\nu{}_{tt} =0$,
 \begin{align}\label{dptds}
\mu {\extd p_t\over\extd s}&=\Gamma^\nu{}_{t\sigma}g^{\sigma\rho}(p_\nu p_\rho-\lambda\Gamma^\tau{}_{\nu\rho}p_\tau)+  \tfrac{\lambda}{2}  \big( \Gamma^\nu{}_{,t}
-   \,g^{\alpha\beta}{}_{,t} \, \Gamma^\nu{}_{\beta\alpha} \big)\,p_\nu \cr
&=\Gamma^\nu{}_{t\nu}g^{\nu\nu}(p_\nu p_\nu-\lambda\Gamma^\tau{}_{\nu\nu}p_\tau)+  \tfrac{\lambda}{2}  \big( \Gamma^\nu{}_{,t}
+2\frac{\dot a}{a}\, g^{\alpha\beta}\, \Gamma^\nu{}_{\beta\alpha} \big)\,p_\nu \cr
  &=\Gamma^\nu{}_{t\nu}g^{\nu\nu}(p_\nu p_\nu-\lambda\Gamma^\tau{}_{\nu\nu}p_\tau)+  \tfrac{\lambda}{2}  \big( \Gamma^\nu{}_{,t}
+2\frac{\dot a}{a}\, \Gamma^\nu\big)\,p_\nu \cr
&=  \frac{\dot a}{a}\,  \sum_{\nu\neq t} g^{\nu\nu}(p_\nu p_\nu-\lambda\Gamma^\tau{}_{\nu\nu}p_\tau)+  \tfrac{\lambda}{2}  \big( \Gamma^\nu{}_{,t}
+2\frac{\dot a}{a}\, \Gamma^\nu\big)\,p_\nu \cr
&=  \frac{\dot a}{a}\,  \sum_{\nu\neq t} g^{\nu\nu} \, p_\nu p_\nu  -  
 \frac{\dot a}{a}\, \lambda \, \Gamma^\tau{} \,p_\tau
 +  \tfrac{\lambda}{2}  \big( \Gamma^\nu{}_{,t}
+2\frac{\dot a}{a}\, \Gamma^\nu\big)\,p_\nu \cr
&=  \frac{\dot a}{a}\,  \sum_{\nu\neq t} g^{\nu\nu} \, p_\nu p_\nu 
 +  \tfrac{\lambda}{2}  \, \Gamma^\nu{}_{,t} \,p_\nu \cr
 &=  \frac{\dot a}{a}\,  \sum_{\nu\neq t} g^{\nu\nu} \, p_\nu p_\nu 
 +  \frac{3\lambda}{2}  \, \frac{\ddot a\,a-\dot a^2}{a^2}\, p_t
 -\lambda\, \frac{\dot a}{a}\, {3 \kappa r^2-2\over a^2 r}\,p_r
 + \lambda\, \frac{\dot a}{a}\,{\cot\theta\over a^2 r^2}\, p_\theta \cr
  &=  \frac{\dot a}{a}\,   \frac{1-\kappa r^2}{a^2} \, p_r p_r 
 +  \frac{3\lambda}{2}  \, \frac{\ddot a\,a-\dot a^2}{a^2}\, p_t
 -\lambda\, \frac{\dot a}{a}\, {2-3 \kappa r^2\over a^2 r}\,p_r
 +  \frac{\dot a}{a^3\, r^2}\, p_{\mathrm{sph}}^2.
 \end{align}
 Equivalently,
 \begin{equation} \mu{\extd P_t\over\extd s}=\frac{\dot a}{a^3}\left( (1-\kappa r^2) p_r p_r 
 +\lambda\,  {2-3 \kappa r^2\over  r}\, p_r
 +  \frac{1}{ r^2}\, p_{\mathrm{sph}}^2\right)={\dot a\over a^3}p^2_\Delta, \end{equation}
 where $p^2_\Delta$ is the operator that acts as $\lambda^2 \Delta$. For the calculation here, we used
\[
\mu\,\frac{\extd f(t)}{\extd s} = -\dot f\, p_t -\frac{\lambda}{2}\, \ddot f - \frac{3\lambda}{2}\,  \frac{\dot a}{a} \, \dot f
\]
for any function $f(t)$ as part of the structure of the differential calculus by previous  methods\cite{BegMa:gen}.

\begin{figure}
\[ \includegraphics[scale=0.65]{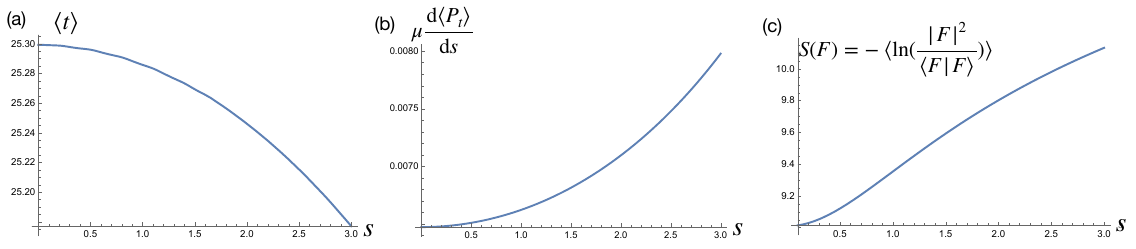}\]
\caption{(a) Expectation value $\<t\>$ for the Gaussian evolution in Figure~\ref{figF} (b)  $\mu {\extd \<P_t\>\over\extd s}$ computed from the RHS of (\ref{dvevPds}) for the same solution. (c) The classical entropy increases during the evolution. \label{figvev}}
\end{figure}

 Note that  since $p^2_{sph}$ involves only angles and angular derivatives, it commutes with the Hamiltonian, which is given by 
 \[ -p^2_{tot}= -(p_t^2+ 3\lambda {\dot a\over a}p_t)+ {1\over a^2}p^2_\Delta \]
 and this obviously also commutes with itself.  Here $-p^2_{tot}$ lands on $\lambda^2\square$ in the Schr\"odinger representation. Thus $p_\phi,p^2_{sph}, p^2_{\square}$ are three constants of motion (useful for finding geodesics in the classical limit where $\lambda=0$ and we treat the $p$'s as real variables), while $P_t$ behaves simply as
\[ \mu{\extd P_t\over\extd s}=\frac{\dot a}{a}( P_t^2-p^2_{tot}).\]
Moreover, on factorisable wavefunctions $\Psi=F(s,t)\psi_\nu(r,\theta,\phi)$ we have
\[\mu {\extd P_t\over\extd s}F= - {\dot a\over a^3}\lambda^2 \nu F\]
and hence combining with (\ref{dvevtds}), we have
\begin{equation}-\mu^2\label{dvevPds} {\extd^2 \<t\>\over\extd s^2} =\mu{\extd \<P_t\>\over\extd s}=\hbar^2\nu{ \int |F|^2\dot a \extd t\over \int |F|^2 a^3\extd t}.\end{equation}
If $F$ is a normalisable stationary state for temporal QM then this should vanish. The $F^\pm_\omega$ in Example~\ref{exH} in the oscillatory regime are not quite normalisable (they are more like plane waves in $t$) but if one regulates both top and bottom by $\int_0^L$ then the right hand side of (\ref{dvevPds}) indeed vanishes as $L\to \infty$.  Likewise, Figure~\ref{figvev} shows both $\<t\>$ and the right hand side of (\ref{dvevPds}) computed for the numerical Gaussian evolution in Figure~\ref{figF}, from which one can verify a reasonable match of the latter (an order of magnitude above the numerical noise) to the left hand side of (\ref{dvevPds}). The expected value drifts downwards about $0.6\%$ over the duration of the plot (which is for $s\in[0,3]$). Note that this is not the evolution of a `physical' wavelike mode, rather we are verifying our formalism and providing proof of concept.  For completeness, in Figure~\ref{figvev}(c), we also computed  the entropy of 
the classical probability density $|F|^2$ for the same evolution, namely
\[ S(F)=-\int {|F|^2\over\<F|F\>} \ln\left({|F|^2\over\<F|F\>}  \right)a^3\extd t.\]
We see that the entropy increases, which is in line with the dispersion evident in Figure~\ref{figF}.

Although we have focussed on the Hubble constant case, there are qualitatively identical results for other positive increasing choices such as $a(t)=({t\over t_0})^{2/3}$, applicable in the matter dominated phase. Here small $t$ has an effective $w=0$ while large $t$ has $w=-{1\over 3}$. We can again find decaying oscillatory stationary modes $F_\omega(t)$ for the temporal system for all real $\omega$ and these decay as $1/t$ and hence are again not normalisable. Likewise, solving the temporal QM equation (\ref{TQM}) for evolution of a Gaussian looks qualitatively as in Figure~\ref{figF} and its evolving expectations and entropy look qualitatively as in Figure~\ref{figvev}.  By contrast, if we let be a decreasing function such as $a(t)=1/t$, one can again solve (\ref{TQMi}) for $F_\omega$ with say value 0 at $t=1$, but these then grow with $t$, and likewise evolution of a Gaussian appears to be unstable as $s$ increases, at least at the numerical level. 

\section{FLRW quantum mechanics with respect to coordinate time}\label{secexp}

KGQM is not ordinary quantum mechanics as it is concerned with quantising geodesic motion and has an external  time parameter $s$ needed for this. We will see in this section that  the tools we have developed can nevertheless be useful also for ordinary quantum mechanics on space with respect to $t$. We take a view on this that does not make `slowly varying' assumptions by considering quantum mechanics as solving the full (not approximated)  KG equation, which is 2nd order. Instead, we can choose a `polarisation' in which we divide the modes into positive and negative energies respect to a natural time variable (in our case $t$) and just chose one of these, so that the 2nd order system with respect to $t$ behaves effectively first oder.  Note that we are not bound to impose a positive or negative energy condition, its role is only for reduction to ordinary quantum mechanics. Moreover, it might be obvious that we should focus on positive energy modes (in the sense of positive frequency) but this will turn out to land us on a conjugate or time-reversed Schr\"odinger's equation in the flat spacetime low energy limit. Exactly the same issue arises in the more conventional route to deriving Schr\"odingers equation from the Klein-Gordon one in which one considers solutions to the latter of the form $e^{-\imath {m_{KG}\over\hbar}t}\psi(t,x)$ with $\psi(t,x)$ slowly varying so that we drop $\ddot\psi$. Then $\square={m_{KG}^2\over\hbar^2}$ reduces to the actual Schr\"odinger equation, but we had to factor out a negative frequency plane wave. 

Next, in order to compare to quantum mechanics,  we extend the Klein-Gordon operator by a potential function $V$ over spacetime. This can be done for  the entire formalism\cite{BegMa:gen}, but here it just means we add $V$ to the wave operator. We assume a potential of the form ${1\over a^2}V(x)$ with a prescribed $t$ dependence in the FLRW coordinates. The separation of variables is then just as before and the only difference is that now $\psi(t,x)=F_\omega(t)\psi_\nu(x)$ has spatial eigenstates obeying
\begin{equation}\label{DeltaV} (\Delta-{2\mu_0\over\hbar^2} V)\psi_\nu=-\nu\psi_\nu, \end{equation}
where we normalise $V$ with a constant $\mu_0$ of mass dimension. This corresponds as in quantum mechanics to energy 
\begin{equation}\label{Enu} E_\nu= {\hbar^2\over 2\mu_0} \nu.\end{equation}
The allowed eigenfunctions (and eigenvalues) are modified according to $V$ but we still suppose that $F_\omega(t)$ obeys the time-independent Schr\"odinger equation (\ref{TQMi}) and multiplying this into $\psi_\nu$ now solves the Klein-Gordon equation-with-potential, with the same mass (\ref{mKG}). The eigenfunctions $\psi_\nu$ depend on $\kappa$ and on the potential, but this is a self-contained problem. For example, solutions in the hyperbolic case for a $1/r$ potential `hydrogen atom' are given via Heun functions. Eigenfunctions and the allowed spectrum (which gets modified by $\kappa$) are also known from a path integral analysis\cite{BIJ}. The more novel part now is about the $F_\omega(t)$, and here we  focus initially on the Hubble constant case where Example~\ref{exH} and previous work\cite{Fer}  already identified suitable positive and negative frequency modes $F^\pm_\omega(t)$.

The first novel feature is that these $F^\pm_\omega$ modes are only oscillatory for
\begin{equation}\label{mH} m_{KG}> {3 \hbar\over 2}H\end{equation}
in terms of the Klein-Gordon mass. So  for solutions of  fixed mass, the behaviour is entirely different if the rate of expansion is too high. For the current epoch, ${3\over 2} H\sim 3.3 \times 10^{-18}$ Hz or $2.2 \times 10^{-42}$ GeV in particle physics units, so well below any nonzero masses in the Standard Model. But working the calculation the other way and using particle physics units, the condition for, say, an electron mass, is that
\[ H< {2\over 3} m_e= 0.34\ {\rm MeV},\]
which already puts is in contradiction with the range
\[ 1 {\rm MeV} < H_{\rm infl} < 10^{10}\ {\rm GeV}\]
considered in most models of inflation\cite{DreXu}. Hence, {\em during inflation, $F^\pm_\omega$ are in their exponential not oscillatory regime}. 

Sticking for the moment with normal small values of $H$, we next consider a formulation of quantum measurement. We take an  initial $\psi(0,x)$ at time $t=0$ and evolve it according to the Klein-Gordon equation (with potential) but keeping only positive energy modes, which we make sense of for each $\nu$ in an expansion of $\psi(0,x)$ in spatial eigenmodes. For each mode we use $F^{\pm,\nu}_\omega(t)$ from Example~\ref{exH} but fixing one of these for our analysis (with the - case corresponding to regular quantum mechanics). Next, we consider $|\psi(t,x)|^2$ as a probability density over space at each time $t$. Here the relevant surface integration is the 4-integral over the volume spanned by a spatial hypersurface  displaced by normal geodesic height $\eps$, divided by $\eps$ and in the limit $\eps\to 0$. In the FLRW case,  this gives an extra factor $a^3(t)$, but this cancels when computing the  expectation of an observable  $\CO$, so that 
\[ \<\CO\>(t)= { \int \extd^3\mu \bar\psi(t,x) \CO \psi(t,x)\over \int \extd^3\mu |\psi(t,x)|^2}, \]
where $\extd^3\mu={r^2\extd r\over\sqrt{1-\kappa r^2}}\sin(\theta)\extd\theta\extd\phi$ is the spatial measure as used elsewhere. 

Next, since the $F^{\pm,\nu}_\omega(t)$ depend on $\nu$, if we expand $\psi(0,x)$ in terms of spatial eigenfunctions $\psi_\nu$ for different $\nu$, the different modes will evolves differently. The same is true in regular quantum mechanics but is now no longer given by a simple phase factor for each eigenstate, since the $F^{\pm,\nu}(t)$ are not simply plane waves in $t$. For example,  
\[ \psi(0,x)=\sum_i c_i\psi_{\nu_i}(x)=\sum_{i} b_i  F^{\pm,\nu_i}_\omega(0)\psi_{\nu_i}(x),\quad \Rightarrow\quad \psi(t,x)=\sum_i b_i F^{\pm,\nu_i}_\omega(t)  \psi_{\nu_i}(x)\]
where we factor out $F^{\pm,\nu_i}_\omega(0)$ from the coefficients in the mode expansion. If we suppose that we have a spatial observable acting as $\CO\psi_{\nu_i}=\sum_iO_{ij}\psi_{\nu_j}$, and that the spatial eigenstates $\{\psi_{\nu_i}\}$ are  orthonormal under the spatial inner product, then 
\[ \<\CO\>(t)={\sum_{ij} \bar b_i b_j  \overline{F_\omega^{\pm,\nu_i}}(t) F_\omega^{\pm,\nu_j}(t)O_{ji} \over \sum_{i}  |b_i|^2  |F_\omega^{\pm,\nu_i}(t)|^2 } . \]
 We first check that  this agrees with quantum mechanics  for small $H$ and small energies, for which we use a different expansion
 \[ F^{\pm,\nu}_\omega(t) =F^{\pm,\nu}_\omega(0)  e^{-{3\over 2}Ht}e^{\pm i \sqrt{\omega^2+\nu}\,t}\left(1\mp \imath{H \nu \over 4 (\omega^2+\nu)^{3\over 2}} \Big(e^{\mp 2\imath \sqrt{\omega^2+\nu} t} - e_{(2)}^{\pm 2\imath \sqrt{\omega^2+\nu} t}\Big)+ O(H^2)\right)\]
 of the bracketed expression in powers of $H$. Here $e_{(2)}^x=1+x+x^2/2$ is the truncated exponential and there is also an assumption $Ht<<1$ in the derivation. We assume  $H<{2\over 3}\omega$ so that we are in the oscillatory regime and if we make the  nonrelativistic low energy assumption $|\nu|<<\omega^2$ then the order $H$ term here is small. Then
 \[ \psi(t,x)\approx e^{-{3\over 2}H t} \sum_i  c_i e^{\pm i \sqrt{\omega^2+\nu}\,t}\psi_{\nu_i}(x)\approx  e^{-{3\over 2}Ht} e^{\pm i \omega t}\sum_i  c_i e^{\pm\imath {E_{
\nu_i}\over\hbar}t}\psi_{\nu_i}(x),\]
where we identify $\mu_0=m_{KG}=\hbar\omega$ to match with  (\ref{Enu}) and (\ref{mKG})  in regular quantum mechanics. The geometric measure for expectation values has an $a^3$ factor  which kills the exponential decay factor both above and below (in particular, the adjusted   length $a^3\sum_{i}  |c_i|^2$ is constant in $t$ in this approximation), but anyway this cancels in the ratio. Thus,
 \[ \<\CO\>(t)\approx {\sum_{ij} \bar c_i c_j e^{\pm\imath{(E_{\nu_j}-E_{\nu_i})\over \hbar} t}  O_{ji} \over \sum_{i}  |c_i|^2 } \]
so that observables behave  as in quantum mechanics for low $H$ and low energies $E_\nu$ provided we choose the $-$ case.  Note that we could have normalised $F^{\pm.\nu}_\omega$ in Example~\ref{exH} to so that $F^{\pm.\nu}_\omega(0)=1$ but we chose not to in order to have good asymptotics for large $t$. However, we also have for $Ht<<1$ that
\[ F^{\pm,\nu}_\omega(0) \approx {}_0F_1\left(1\pm\imath{\omega'\over H}, -\frac{\nu}{4 a_0^2 H^2}\right)\approx 1\pm\imath  \frac{\nu}{4 a_0^2 H \omega}+ O(\nu^2),\]
where in the second step it is convenient to assume $\omega'=\sqrt{\omega^2-{9 H^2\over 4}}\approx\omega$ or $H<<\omega$ (i.e., well inside the oscillatory regime). 
Hence,  $F^{\pm,\nu}_\omega(0)\approx 1$ would require a stronger `very low energy assumption' $|\nu|<< 4 a_0^2 H\omega$ which, certainly in the current epoch, would not apply for normal energies.

 \begin{figure}
\[ \includegraphics[scale=.75]{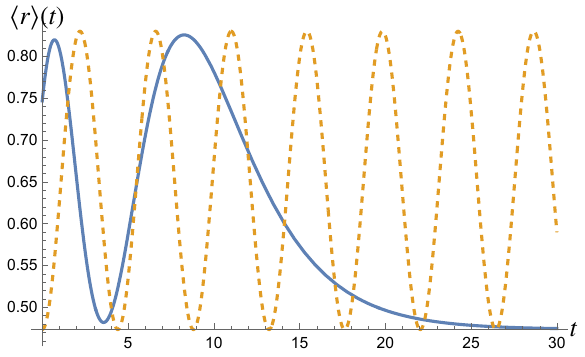}\]
\caption{Expected value $\< r \>(t)$ in KG solution/KGQM stationary state $\psi=F^{-,8}_{\omega}(t)\psi_{3,0}(r)+ F^{-,24}_\omega(t)\psi_{5,0}(r)$ evolving a sum of  $l=0$ cosmological atom states for levels $n=3,5$. The dashed line is the corresponding behaviour in regular quantum mechanics oscillating between the  expected value in states $\psi_{3,0}\pm\psi_{5,0}$. Plots are for $\kappa=a_0=1$, $H=0.08$, $\omega=4$. 
\label{figO}}
\end{figure}

For any finite $H>0$ and larger $t$, however, both numerator and denominator of $\<\CO\>(t)$ depend on $t$ but tend to a constant value if we include the $a^3$ factor in each (and the same for the ratio even without this factor). This follows from the large $t$ expansion (\ref{oscappx}), where $F^{\pm,\nu}_\omega$ become independent of $\nu$,  and applies to any spatial operator $\CO$. For the matrix case discussed above, it would be
\[  \<\CO\>(t)\sim  {\sum_{ij} \bar b_i b_j O_{ji} \over \sum_{i}  |b_i|^2 }\]
for large $t$, but we are not limited to this case.  A concrete example  $\CO=r$ is given in Figure~\ref{figO}, where we plot $\<r\>(t)$ for $\psi$ a linear combination of cosmological atom states $\psi_{3,0}(r)$ and $\psi_{5,0}(r)$, cf  Figure~\ref{fig1}, evolved with the relevant $F^{-,\nu_i}_\omega(t)$. This initially oscillates (as in regular quantum mechanics) between $\<r\>$ in the sum and difference states ${\psi_{3,0}\pm \psi_{5,0}}$, before tending to the former. The plots of $\<r^\alpha\>$  for other powers, including negative ones, behave in the same way.  In effect, the expected quantum mechanical  behaviour is present for small $t$ but gets `washed out' by the expansion. This happens for 
 \[ t>> {1\over H}\ln({ \sqrt{2 |E_\nu|}\over  a_0 \sqrt{m_{KG}}}) \]
from (\ref{tc}) and (\ref{Enu}) after we identify $\mu_0$ with the mass of the Klein-Gordon field. We can also write this as
 \[ a(t)<< \sqrt{2 |E_\nu|\over m_e}\]
using, say, the mass of the electron.  This is not relevant in the current epoch, but could be relevant when the Universe was smaller. For example, using the highest energy of a hydrogen atom with $|E_\nu|\approx 14$ eV, we need $a(t)<<0.75\%$ (compared to its current value usually taken to be 1). 
 
\begin{figure}
\[ \includegraphics[scale=.7]{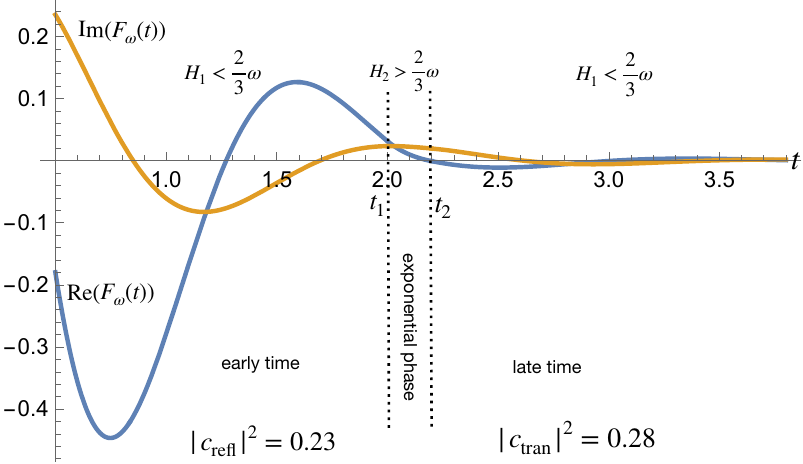}\]
\caption{Solution for $F_\omega(t)$ covering an exponential period $t\in (2,2.2)$ where the Hubble constant jumps to above the bound (\ref{mH}) for the oscillatory regime (as could happen during inflation). The imaginary part looks broadly similar. The plot is for $a_0=\nu=1$, $\omega=4$,  $H_1=1$, $H_2=2.7$.\label{figtrans}}
\end{figure}

Finally, returning to inflation, suppose we have a period of slow expansion with $H=H_1$ for $t<t_1$ such that the $F^{\pm,\nu}_\omega$ are in the oscillatory regime, followed by a period of higher  expansion $t_1<t<t_2$ with $H=H_2$ putting us in the exponential regime (as discussed above), and then maybe another period $t>t_2$ of slow expansion with $H=H_1$. This is not a single Hubble constant as discussed above but we can still solve for solutions $F_\omega$ of the time-independent temporal Schr\"odinger equation (\ref{TQMi}) for this composite  $a(t)$. This can be done in first approximation by piecewise matching exactly as in elementary quantum mechanics in 1-dimension in the presence of   a potential barrier, just with $x$ replaced by $t$. Thus, for times that correspond to each value of $H$ we consider in principle both $F^{\pm,\nu}_\omega(t)$ modes and require continuity of $F_\omega$ and its derivatives at the transition points. To have match regular quantum mechanics as much as possible, we fix `incoming' $F^{-,\nu}_\omega(t)$ with coefficient 1 for small $t$ and the `outgoing'  $F^{-,\nu}_\omega(t)=0$. The other coefficients are free, so 
\[ F_\omega(t)=\begin{cases} F_\omega^{-,\nu}(t)+  c_{\rm refl} F_\omega^{+,\nu} (t)& t\le t_1,\\ a F_\omega^{+,\nu}(t)+ b F_\omega^{-,\nu}(t) &  t_1\le t \le t_2,\\ 
c_{\rm tran} F^{-,\nu}_\omega(t) & t\ge t_2\end{cases}\]
with four complex parameters $a,b,c_{\rm refl}, c_{\rm tran}$, to be determined by the matching at $t_1,t_2$. To have a solution, one is generically forced to have a nontrivial reflection coefficient $c_{\rm refl}$ and there is a transmission coefficient $c_{\rm tran}$. Technically, we should also smooth out the double derivative as our theory is second order, but ignoring this, we arrive at a particular $F_\omega(t)$. The general formulae for the above are not very illuminating to write down, but we show an illustrative example in Figure~\ref{figtrans}. 

The necessity of $F^{+,\nu}_\omega$ `reflected modes' here takes us beyond regular quantum mechanics at early time, but makes sense in the more general second order theory. Then there is an exponential period, and then at late times we have only $F^{-,\nu}_\omega$ modes and see something that looks more like regular quantum mechanics. The term `reflection' as well as `transmission' and the `tunnelling probability' (which in ordinary quantum mechanics would be $|c_{\rm tran}|^2$) make sense in the context of temporal quantum mechanics on the time axis and provide a picture of the above process. Thus, we see that we can use these methods to construct at least piecewise solutions of the Klein-Gordon equation on the FLRW background and  can also use them for the relevant time evolution in the quantum-mechanics-like interpretation here with respect to $t$. These KG solutions are given by multiplying $F_\omega(t)$ solved for $\nu$ by any spatial eigenfunction of the spatial system (\ref{DeltaV}) with eigenvalue $-\nu$. While $\nu>0$ for the cosmological atom modes, one can equally well use the relevant potential $V$ for actual hydrogen atom modes when $\kappa=0$, where $\nu<0$, and similar modes when $\kappa<0$. The solutions as in Figure~\ref{figtrans} look much the same when solved for $\nu=-1$ and multiplying by $F_\omega(t)$ then evolves such atomic spatial modes into solutions of the KG equations through the period of inflation.  

\section{Concluding remarks}\label{secconc}

We explored the use of a mathematical tool involving a hypothetical external `observer' with time $s$ able to see, at the classical level, each world-line evolving by this amount of its own proper time. This is not fundamentally different from other types of flow in geometry, notably if $\hbar=0$ then (\ref{heis1})-(\ref{heis2}) reduce to a flow on the cotangent bundle as used, for example, in \cite{Cha}. The new feature is that these are now quantised to operator equations on the Heisenberg algebra. Corresponding to this, we looked at quantum mechanics with respect to this external observer's time $s$ and with the Klein-Gordon operator generating the evolution. This provides a new context in which we can apply tools and ideas from quantum mechanics  covariantly and in a coordinate-independent manner\cite{BegMa:gen}. 

Applications to black holes in that work and to FLRW cosmology in the present work show that this novel point of view leads at minimum to interesting stationary states, i.e. solutions of the Klein-Gordon equations,  that could respectively be called `gravatoms' and `cosmological atoms'. In both cases the KG solutions have a separated form, where the $t$-dependence factors out as a plane wave $e^{-\imath\omega t}$ in the black hole case and as more complicated functions $F_\omega(t)$ in the FLRW case. Focussing on such solutions and their associated boundary conditions breaks the diffeomorphism invariance but in both cases these are standard coordinates with a known relation of Schwarzschild time and cosmological time to other physics. Our cosmological atom solutions provide, in particular,  precise meaning to the idea that if the Universe is bounded then there should be harmonic modes spanning it. We found in Section~\ref{secatom} for $\kappa>0$ that these are labelled by $n=2,3,\cdots$  and certain allowed angular quantum numbers $(l,m)$ depending on $n$. 

We also studied in Section~\ref{secexp} how exactly our approach is relevant to viewing solutions of the Klein-Gordon equations as quantum mechanics on space with respect to coordinate time $t$, with non-oscillatory $F_\omega(t)$ when the Klein-Gordon mass is below ${3\hbar\over 2}H$, which is typically the case during inflation.  Another potentially physical effect relevant to inflation  was that a sudden change in $a(t)$ generates a reflected wave in the  behaviour of $F_\omega(t)$, and we also saw the existence of other effects  at times where $a(t)$ is small. These new phenomena should be looked at further in the context of evolving spatial modes (such as an actual atom) through a period of inflation as a solution of the KG equations. Inflation itself has a long history initially   motivated by the elimination of magnetic monopoles\cite{Gut}, but can also be motivated by possible quantum gravity effects in the early universe. Moreover,  temporal quantum mechanics (\ref{TQM}) for constant $H$ has a Liouville potential and could be interesting to approach by other methods\cite{Jac,Kob}, also using Bessel functions.

For non-stationary states of KGQM, where wave functions evolve with $s$, we  found that due to a different nature of the metric, the separation of variables leads in the FLRW case to an effective 1-dimensional quantum mechanics over the $t$-axis (so this plays the role usually played by the $x$-axis) and with the expansion factor $a(t)$ as both a potential and a measure of integration. We called this factor of the wave function $F(s,t)$ and showed that there is a quantum mechanics-like theory with respect to $s$, including Ehrenfest theorems coming out of the operator-algebraic `Heisenberg' picture.  Notably, we found an expression  (\ref{dvevPds}) for the acceleration of $\<t\>$ with respect to $s$. In the classical limit, this necessarily behaves as expected for $s$ the proper time of a single geodesic (as for flat spacetime in \cite{BegMa:geo}). Applying such new operator tools to cosmological problems would be another important direction for further work.

Another area for further study would be to extend our focus on factorised states $F(s,t)\psi_\nu(r,\theta,\phi)$ to the general case
\[\psi_s(t,x)= \sum_{\nu} F_\nu(s,t)\psi_\nu(r,\theta,\phi)\]
viewed as entangling the temporal and spatial systems. Here,  we sum (or integrate) over different eigenstates of the spatial Laplacian $\Delta$, each with their own $F_\nu$ factor.  From this point of view, one can trace out the spatial sector to create thermal and other mixed states or density matrices in the temporal quantum mechanics. The natural choice here would be expectation values of the form
\[ \<a\>=\sum_{\nu}e^{-\beta\nu} \<F_\nu |a | F_\nu\>\]
for an ensemble of normalised states $\{F_\nu\}$ and an observable $a$ in the temporal sector. At least for a static metric, one could similarly trace out the temporal sector to create density matrices in the spatial sector of the theory.

\section*{Acknowledgments}
\noindent 
SM was supported by  Leverhulme Trust project grant RPG-2024-177

\section*{Declarations}

\medskip
\noindent{\bf Data availability:} Data sharing is not applicable as no data sets were generated or analysed during the current study.

\medskip
\noindent{\bf Conflict of Interest:} On behalf of all authors, the corresponding author states that there is no conflict of interest.

\end{document}